# Quasicrystals in the molecular–dynamic model of pure aluminum

A. S. Prokhoda[*] and A. M. Ovrutsky

*Dnepropetrovsk National University, Dnepropetrovsk, Ukraine*

Growth and structures of crystals in the model of Al obtained in results of isothermal annealing after quick cooling to certain temperatures are studied by the method of molecular dynamics applying the known potential of EAM type. The growing nanocrystals have not fcc crystal structure in spite of it stable when setting in initial condition. We have determined two types of crystallization centers with hexagonal and rhombic crystalline lattices. Both lattices are built from hexagonal clusters of 15 atoms (the Frank–Kasper type). In many places, small crystalline chips with these lattices coincide one with another forming pictures typical for quasicrystals of dodecagonal type; and translation symmetry in direction perpendicular to the sections with hexagons takes place. The tilling picture for the obtained structure is constructed and it is compared with the tilling picture for the known quasicrystals of dodecagonal type. The nature of forming of such quasicrystals is considered including the question on additional Laue patterns of twelfth order.

PACS numbers: 61.20.Ja, 61.44.Br, 64.60.De, 63.20.Ry, 81.10.Aj.

Keywords: computer simulation; quasicrystals; nucleation; growth; structure; metals.

## 1. Introduction

Comparatively large atomic systems can be simulated with the help of modern personal computer with modern graphic cards (power GPUs) when using the program packages for parallel computations, for instance LAMMPS. Such computer systems allow obtaining results of simulations for the models from $10^5$ atoms during periods of 1÷10 ns in several workdays. Many physical properties of metals and alloys are featured correctly when using ab-initio calculation or up-to-date EAM or MEAM potentials. Therefore, the structures of solidificated systems obtained in results of simulations can correspond to real structures.

______________________________

*Email: a-prokhoda@mail.ru

Five-fold Laue patterns were discovered at first in metals with fcc crystalline lattice, crystals of which were periodically twinned with rotation of the crystalline lattice in 72° (5 sectors with {220} planes in the section perpendicular to 5-fold axis). These structures were studied in many great works, including the studies fulfilled in last years [1-4].

The appeal of fivefold symmetry was tremendously encouraged with the disclosure of icosahedral quasicrystals [5-7] and with the invention of the quasilattice concept to describe these structures basing on local icosahedral packing of atoms contained in tetrahedrally close-packed intermetallic compounds [8-10].

More than 100 different alloys are known now as those, in which quasicrystals can arise. A good few from them are the alloys containing aluminum. Quasicrystalline materials with icosahedral quasicrystals show quasiperiodicity in all three dimensions. The other classes of quasicrystals – octagonal, decagonal, and dodecagonal – are quasiperiodic in two directions, in the quasiperiodic plane, and periodic in the perpendicular to, the quasiperiodic plane direction. Lists of alloys having quasicrystalline structures of icosahedral, octagonal, decagonal, and dodecagonal symmetry are collected in Refs. [11,12].

When studying formation of fivefold twin structures by simulations in Ref. [4], we had applied for aluminum the potential from Ref. [13], as the more known potential from Ref. [14] did not provide crystallization during comparatively long time in spite of crystals with fcc-lattice set in initial condition have normal kinetics of growth. In these work, we used for simulations of sufficiently large models the potential from Ref. [14]. Its pair part has two gently sloping minima. Most likely, this is a cause of variety of simulated structures.

## 2. Details of the simulation

The MD simulations were fulfilled using the LAMMPS codes (large-scale atomic-molecular massively parallel simulator) for parallel computing with timestep of 2 fs (the Verlet algorithm in velocity form, *NVT* ensemble and Berendsen thermostat). Spherical models with free surfaces (of 131072 atoms) were obtained in results of melting of the first-set crystals and quick cooling of liquids to the chosen temperature of annealing. The melting temperature $T_m$ of the crystal with the fcc lattice set in initial condition was determined. For the potential from Ref. [14], $T_m = 870 \pm 5$ K.

Identifications of clusters were fulfilled in results of running of the LAMMPS codes (compute cna/atom and pe/atom) or using the programs for visualization: VMD and OVITO. When visualizing the results of simulations, we set colors for atom dependently on cluster type, i.e. on their neighbor surroundings.

## 3. Results and their discussion

We found that in the supercooled liquid state, our model has many icosahedral clusters As one can see from Fig. 1,a, LAMMPS determines many icosahedral clusters (their central atoms are red) not only in liquid, but also in crystallization centers (they are seen at certain orientation, Fig.1,b), especially on their outsides. And there are many hexagonal clusters in the shown sections (see Fig. 1,a), which are ordered in nanocrystals. At the temperature 500 K in liquid state, LAMMPS marks ~ 8% from the full number of atoms as the centers of icosahedral cluster. If take into account the nearest neighbors of these atoms, we can say that the majority of atoms belongs to the clusters. Therefore, the nucleation of crystal phase is very difficult. The waiting time for nucleation at this temperature was roughly 1 ns.

Figure 1 shows crystallization. There are many ordered places. Some of them are nanocrystals from long hexagonal tubes placed into the square lattice (with full periodicity along the tubes). Before crystallization, we see in liquid sufficiently many nuclei from hexagonal clusters in the form of rhombs. However, the major of ordered places are constructed from small crystalline chips (small nanocrystal) having strongly square or rhombic shapes in one section. This is a typical picture for quasicrystals of dodecagonal type; we will name their below as "our quasicrystals". It is most important that two types of crystals, from which our quasicrystals consist, may exist separately.

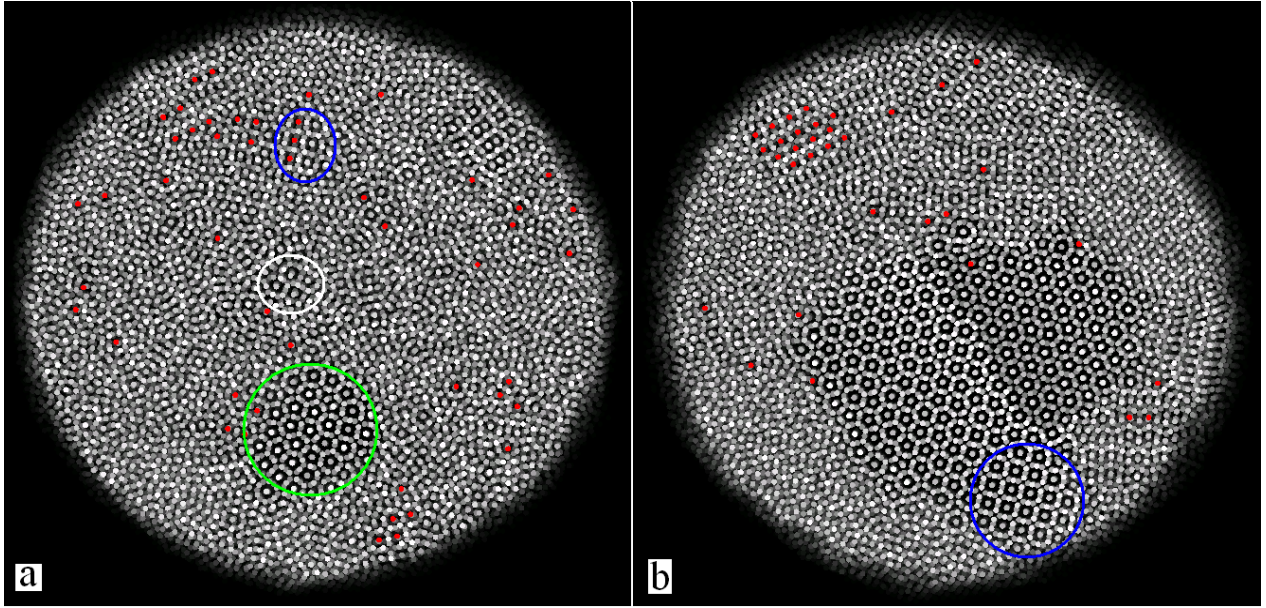

**Fig. 1. Snapshots of the model sections during annealing, temperature *T*=500 K; time of annealing is 1.6 and 2.2 ns for figures a and b, accordingly; centers of icosahedral clusters are red; a white circle shows the nucleus of the hexagonal phase; blue circles show nanocrystals with a rhombic crystal lattice; green circle in Fig. 1, a shows a quasicrystal – it is large in Fig. 1, b.**

We have determined exactly the crystalline lattices of two crystalline phases. The hexagonal phase has an elementary cell from 19 atoms (Fig. 2,a). The large crystal with this structure set in initial condition is stable up to 300 K (P6mm type of crystalline lattice). At higher temperatures, the reconstruction into other structures (another crystalline phase and quasicrystals) takes place. Another phase has the tetragonal close to cubical elementary cell from 20 atoms (Fig. 2,b) (P222 rhombic type of crystalline lattice). This phase is very stable. Its temperature of melting ($T_m$ = 900±5 K) is higher than the temperature of melting of the fcc-phase. In both lattices, the hexagonal clusters form long tubes in the direction perpendicular to the plane of hexagons. Our quasicrystals are also stable. Most big from them was melted at the temperature ~865±5 K.

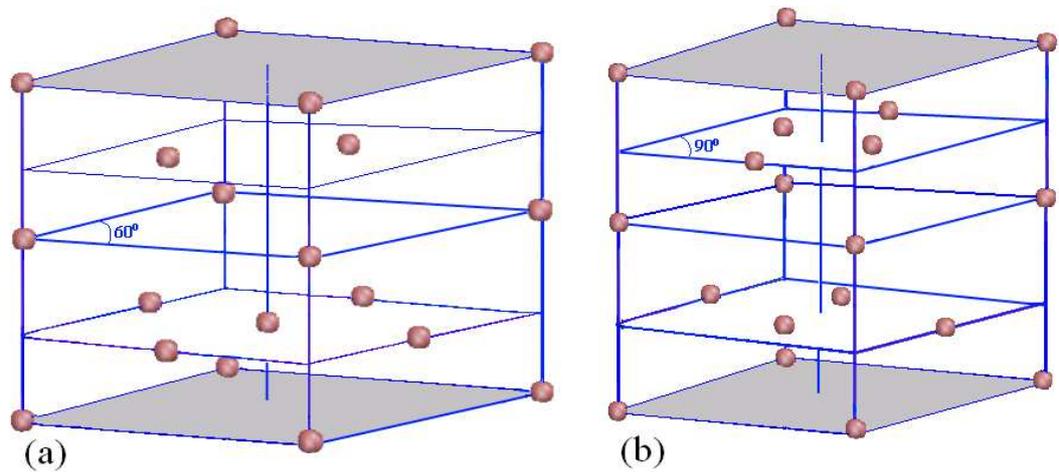

**Fig. 2.** The cells of two lattices that correspond to two crystal phases,
(a) the cell of hexagonal lattice (the cell parameters: a=b= 5.22 Å, c=5.12 Å);
(b) the tetragonal cell (the cell parameters: a=b= 5.15 Å, c=5.12 Å).

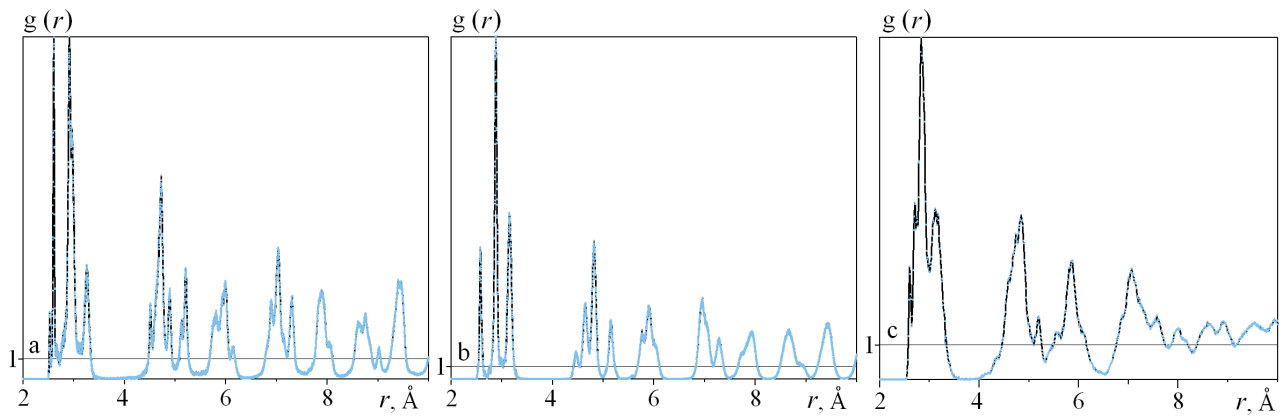

**Fig. 3.** Radial pair distribution functions for phases
with hexagonal (a) and rhombic (b) crystal lattice and for the quasicrystal area (c).

Fig. 3 shows radial pair distribution functions (RPDFs) calculated for separate phases with hexagonal and rhombic crystal lattice and for the quasicrystal area. All RPDFs have divided first peaks. In Fig. 3,a for the hexagonal phase, there are 3 large peaks at 2.61, 2.94 and 3.26 Å and pre-peak at 2.56 Å (c/2). Rhombic phase has peaks at 2.58 (c/2 ?), 2.89, and 3.16 Å. Close to them are the peaks calculated for the quasicrystal area: 2.59, 2.83, and 3.14 Å; these peaks are less divided. In addition, one new peaks at 2.72 Å appears in Fig. 3,c. Most likely, it is connected with many contacts between crystal chips of two phases (see below).

Fig. 4 shows (from two points of viewing) tubes from hexagonal clusters from which the crystalline lattices are built (with the help of VMD software). For these figures, real coordinates of atom were used. The main features of these clusters are in two points: the central atoms are not placed in the planes of hexagons; consecutive hexagons are turned one relative another with the angle 30°. Distances between near are atoms are close for the both crystal lattices (see above); there are smallest distances in 2.53 Å in the hexagonal phase. Small distances are, between atoms on the axes of the hexagonal tubes 2.59 Å. These atoms have 14 nearest neighbors, i.e. they are centers of the clusters of Frank-Kasper type, which are formed from two decahedral groups of atoms (see Fig. 5).

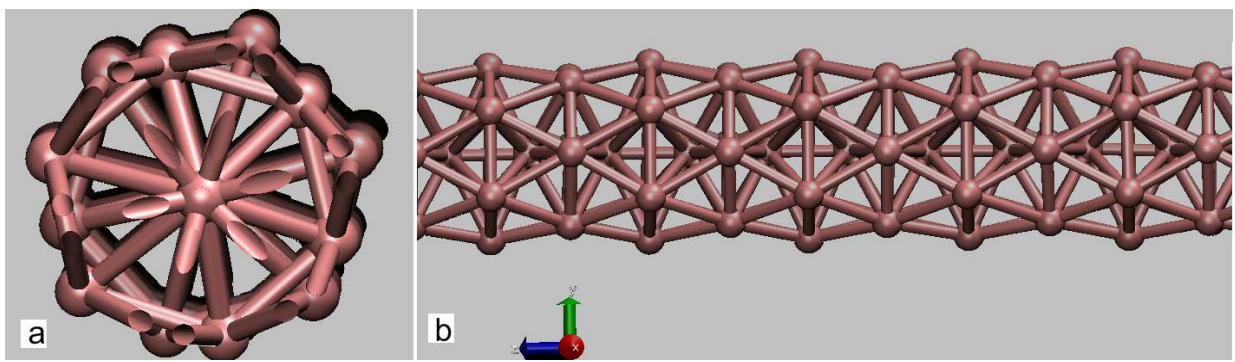

**Fig. 4. Grouping of dodecagonal clusters into a tube; two points of viewing.**

Fig. 5 shows the structure of one bi-dodecagonal cluster (three point of viewing). Groups of dodecagonal clusters, which form tubes, can be considered as molecules. Different packing of these molecules gives different crystalline lattices or our quasicrystals. Two consecutive hexagons from consecutive sections look like a bi-dodecagon with 14 atoms around the central atom (the cluster of Frank-Kasper type). Therefore, one can mistakenly see a symmetry axis of twelve orders (Fig. 1, 4 and 5). Of course, he will see an orientation ordering of rhombs and squares in quasicrystals. Moreover he can build the tilling picture (read about types of tilling in [15]) by joining of the central atoms of hexagonal clusters (such as it is shown in the Fig. 6,a) that shows filling of all area by two types of tiles – squares and triangles. It will be similar to typical tilling pictures for the quasicrystals of dodecagonal type shown in Fig. 6,b. (the square and hexagonal tiles are typical for structures of the intermetallic compounds $Cr_3Si$ and $Zr_4Al_3$, that well-known in the literature [16]). But in our opinion, the picture showing outlines of the crystalline chips of two phas-es (see Fig. 6,c) is more informational, as it responds to physical nature of such structures.

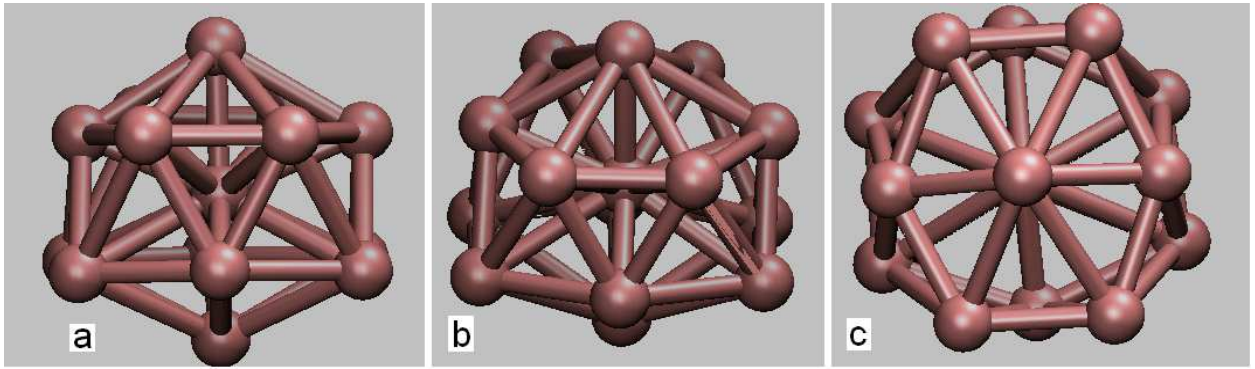

**Fig. 5.** The structure of one bi-dodecagonal cluster.

Is the structure shown in Fig. 6,a the structure of quasicrystal? Yes, it is typical for quasicrystals! Will the Laue patterns from such structure contain the patterns of 12-fold symmetry? Yes, because of consecutive contacts of rhombic and square crystalline chips often lead turning of their orientation in 30°. Fig. 6,c shows the recognized quasicrystalline structure of alloy $(Ta,V)_{1.6}Te$. We marked out by solid and dotted white lines the square and rhombic crystalline chips. The chip from 6 rhombs in the section contains many hundreds of atoms. If contacts between rhombs and squares happen trough the planes of constant directions, their orientation does not change. However, joining of the rhombs to another side of the squares changes its orientation in 30° (from right and in the bottom in Fig. 6, c). Thus, the Laue patterns of 12 order take place due to orientation dependence of two contacting crystalline chips of different phases. Could say, the quasicrystals of dodecagonal types are similar to twinned structures from two phases, for instance to structures with alternate fcc and hcp-phases. It is known that quasicrystals of dodecagonal type consist from different structure units. As rules of stacking of resembling crystalline lattices are the main condition of forming dodecagonal quasicrystals so that a wide variety of packing of small crystals do not damage an orientation order. A definition "two-phase fractal" is a good name for such structures.

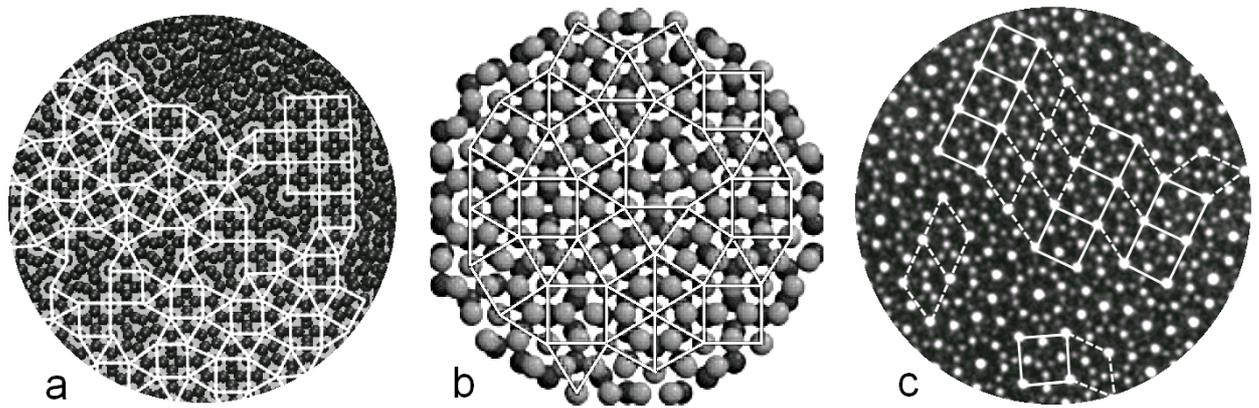

**Fig. 6.** The quasicrystal structures, tilling pictures, and contacts of crystalline chips in the quasicrystal, a – in the simulated model of pure aluminum; b – the typical quasicrystalline structure from long molecules [17]; c – the Google image of the quasicrystalline structure in the alloy $(Ta,V)_{1.6}Te$ obtained by HRTEM.

Placing of peaks in calculated RPDFs (Fig. 3) argues also for two-phase structure of our quasicrystals (the main first peaks are close to those, which are correspond to hexagonal and rhombic phases). And only one addition peak (2.72 Å) in Fig. 3,c argues in favor of contacts of one type between two phases.

Else, one question is interesting for consideration. Are there defects of stacking in quasicrystalline structures, and how they influence the Laue patterns? Of course they are. One can see them in Figs 6,a and 6,c. However, the regular orientation dependence of two phases stacking allows passing around them during crystallization without damage of two major orientations of structure elements.

Dzugutov [18] first had described a molecular dynamics simulation where he had obtained a dodecagonal quasicrystal by a freezing simple monatomic liquid (periodical conditions for comparatively small main cell). Its diffraction pattern closely resembled that of the dodecagonal quasicrystalline phase formed in $V_3Ni_2$ and $V_{15}Ni_{10}Si_{75}$ alloys. Roth and Denton [19] had studied stability of structures which could be formed in the model with Dzugutov's pair potential at different pressures and rates of cooling. They found that bcc crystal structure is equilibrium at low pressures and the σ-phase that is one from tcp-variants of Frank-Kasper phases [16] has a sufficiently low potential energy in wide range of pressures. The stability is lowest for the purely triangular Z-phase (only triangular tiles). More complicated crystalline phases, approximants, and the quasicrystals all contain mixtures of square and triangular tiles in different arrangements. These structures are all inferior to the σ-phase since they must contain larger conglomerates of triangles.

Thus, our results in many respects are similar with those that obtained via Dzugutov potential. The hexagonal phase (only triangular tiles) is the less stable in our case. However, our results obtained for the concrete metal – aluminum that is a basic element of many important intermetallic compounds and great many quasicrystals. Then, we simulated sufficiently large system with free surfaces. Therefore, our model approached to equilibrium during isothermal annealing (at least, to the equilibrium volume).

## 4. Conclusions

The solidificated models of aluminum contain different crystalline structures. There are small nanocrystals with the rhombic crystalline lattice, nuclei with the hexagonal crystalline lattice, and "quasicrystals" of dodecagonal type. Both crystalline lattices consist from hexagonal clusters, which packed in long ordered tubes.

Quasicrystals obtained consist from crystalline chips of two crystalline phases stacking one with another by the regular way with conservation of their orientation or with change of their orientation in 30° or 90°. Thus, the certain orientation order is kept during crystallization. Due existing of two preferred orientations of the crystalline chips in quasicrystals, the Laue patterns from such structures show the existence of symmetry axes of 12 order.

In our opinion, a name "two-phase fractal" is a more exact definition of quasicrystals of dodecagonal type, as stacking rules of different crystalline chips are the main condition of their forming.